\documentclass[a4paper,aps,pre,twocolumn,groupedaddress,showkeys,showpacs,floats,floatats,floatfix]{revtex4-1}
\usepackage[utf8]{inputenc}
\usepackage[english]{babel}
\usepackage{graphicx}
\usepackage{dcolumn} 
\usepackage{bm}
\usepackage{natbib}
\usepackage{latexsym}
\usepackage{mathrsfs}
\usepackage{amssymb}
\usepackage{amsmath}
\usepackage{amscd}
\usepackage{color}
\usepackage{pifont}
\usepackage{pstricks,pst-node,pst-text,pst-3d}
\usepackage{verbatim}
\usepackage{ulem}
\usepackage[T1]{fontenc}
\usepackage{hyperref}
\hypersetup{
  colorlinks   = true, 
  urlcolor     = magenta, 
  linkcolor    = red, 
  citecolor    = blue 
}
\newrgbcolor{Red}{1.0 0.0 1.0}
\begin{document}
\title{Controlling intermediate dynamics in a family of quadratic maps}
\author{Rafael M.~da Silva$^{1}$, Cesar Manchein$^{2}$, and Marcus W.~Beims$^{1}$}
\affiliation{$^1$Departamento de F\'\i sica, Universidade Federal do Paran\'a,
         81531-980 Curitiba, PR, Brazil}
\affiliation{$^2$Departamento de F\'\i sica, Universidade do Estado de
  Santa Catarina, 89219-710 Joinville, SC, Brazil} 
\date{\today}
%
\begin{abstract}
The intermediate dynamics of composed one-dimensional maps is
used to multiply attractors in phase space and create multiple 
independent bifurcation diagrams which can split apart. Results 
are shown for the composition of $k-$paradigmatic quadratic maps with 
distinct values of parameters generating $k-$independent
bifurcation diagrams with corresponding $k$ orbital points. For specific 
conditions, the basic mechanism for creating the shifted diagrams is the 
prohibition of period doubling bifurcations transformed in saddle-node 
bifurcations.  
\end{abstract}
%
\pacs{05.45.Ac,05.45.Pq}
\keywords{Quadratic map, chaos, bifurcation.} 
\maketitle

{\bf
In generic dynamical systems it is very useful to determine the
right parameter combination for which regular or chaotic behavior takes
place. It is more interesting to know or control the robustness 
of the dynamics under certain time parametric changes. With this aim in
mind we present in this work a recipe to perform time parametric
changes in order to control the intermediate dynamics of composed
maps. By using the composition of a family of quadratic maps, it is
possible to generate multiple shifted similar independent bifurcation
diagrams and by consequence the same number of attractors in phase
space. We show that for specific parameter combinations occurs a 
prohibition of period doubling bifurcations and the appearance of extra-stable 
motion. As unimodal maps follow universal bifurcation rules we believe that the 
proposed method is generic and may be extended to ordinary higher-dimensional 
problems involving nonlinear behaviours. 
}

\section{Introduction}
\label{intro}
Nonlinear dynamical systems are one of the most important tools to
model a large number of physical systems in nature, ranging from
biological populations, coupled networks, market crisis, brain 
dynamics, chemical systems, laser physics, granular dynamics, normal
and anomalous transport, extreme events, weather forecast, among many
others. The parameter values in such physical systems deliberate
the underlying complex dynamics. One of the greatest challenge in 
nonlinear models is to identify what are the correct parameter
values which leaves to the desired dynamics (or avoid it) and if 
these values follow some kind of universal rule.

Once the above mentioned desired parameter combination is
found, we ask what is the robustness of the dynamics under time
parametric changes. This is of relevance for practical purposes
where nonlinear models describe realistic problems. In this context,
weak parametric perturbations were used to control the chaotic
motion \cite{loskutov93,mirus99,eno03}, to mention a few. In the
present work we use the concept of parametric changes to 
control the intermediate dynamics (defined in Sec.~\ref{inter}) of composed   
maps. Our goal is to analyse the whole change in the dynamics,
including regular  and chaotic motions. We show that using the
composition of a family of quadratic maps \cite{ott-book}, it is
possible to generate multiple shifted independent  
bifurcation diagrams and the same number of attractors in phase space. 
For specific cases discussed later, we show that the mechanism of
shifted bifurcation diagrams and prohibition of period doubling
bifurcations are responsible for the appearance of extra-stable
motion. An analogous mechanism for just two shifted bifurcation
diagrams was revealed many years ago \cite{eno03} in the context of
taming chaos in continuous-time systems under weak harmonic
perturbation (see also Ref.~\cite{Qu1995}). In the context of chaos
suppression, a similar method was used in the particular case of
duplication in the factorization of two quadratic coupled  maps
\cite{kurths96} and one-dimensional maps were also used for
applications of feedback and non-feedback control techniques
\cite{sanju93,Buchner2000,Nandi2005}. {Periodic perturbations 
can also be applied to determine the role of extreme orbits in 
the organization of periodic regions in the parameter space of 
one dimensional maps \cite{leonel16}}.

Our results are not restricted to weak parametric perturbations and 
we show cases of duplication, triplication, quadruplication and 
quintuplication of bifurcation diagrams and corresponding attractors. 
In addition, since we use the composition of unimodal maps, which have 
only one critical point and display very similar dynamical behaviour, 
the proposed method is generic and can be applied to ordinary higher-dimensional 
problems involving nonlinear behaviours. 

The remainder of this paper is organized as follows. In
Sec.~\ref{inter} the general concept of intermediate dynamics is given
in terms of {the composition of} maps. Section~\ref{QM} is devoted to
discuss the main mechanism which induces the shift of bifurcations
diagrams in the quadratic map, and consequently the proliferation of
periodic attractors in its phase space. In addition, the application
to du, tri, quadru and quintuplicate  periodic attractors is
discussed in the same section. Finally, in Sec.~\ref{conclusions}
we summarize our results indicating some of their possible
implications. 

\section{Intermediate dynamics}
\label{inter}

The key idea to generate multiple attractors is to control the
dynamics of {\it intermediate} variables of {{\it composed}} maps. 
{To explain this in details consider a one-dimensional discrete 
map given by
\begin{equation}
\label{f}
x_{n+1} = f(x_n,\alpha),
\end{equation}
with $n=0,1,2,\ldots$ being the discrete times, $x_n$ the state of the system
at time $n$ whose time evolution is described by the function $f$ (usually 
nonlinear) and $\alpha$ representing all involved parameter. {Now, we
construct a composed map obtained by applying the map (\ref{f}) $k$-times
using distinct parameters at each iteration, namely
\begin{equation}
\label{fc}
x^{(c)}_{n+k} = f(x_{n+k-1},\alpha_k)\circ\ldots\circ 
           f(x_{n+1},\alpha_2)\circ f(x_n^{(c)},\alpha_1),
\end{equation}
where parameters $\alpha_1,\alpha_2,\ldots\alpha_k$ follow a specific 
protocol defined later. In this work the superscript $(c)$ indicates 
quantities related to the composed map. All states from $x_{n+1}$ to 
$x_{n+k-1}$ are called {\it intermediate states}.}
The dynamics of each intermediate state can be controlled and manipulated 
independently. {As we will see, $k$-independent intermediate 
dynamics (chaotic or regular) are generated. It could also be possible
to use distinct functions $f$ at each intermediate dynamics, but such 
case will not be discussed here}. Another approach would be to define new 
variables for each intermediate dynamics, which corresponds to consider a 
higher-dimensional  complex system. 

The concept of intermediate dynamics has been used in open 
chaotic flows to describe the particle's position after a half period 
\cite{review17} and in chemical oscillations where the concentrations of 
intermediates reactants were allowed to vary \cite{noyes76}.}
Not unintentional, the composition used in this work is
analogous to  external controlling forces and can promptly be
implemented in experiments. Some of the related references are given 
in Sec.~\ref{intro}. 

\section{Composed quadratic maps}  
\label{QM}
\subsection{Duplication ($k=2$)}
\begin{widetext}
$\quad$
\begin{figure}[!h]
  \centering
  \includegraphics*[width=0.98\columnwidth]{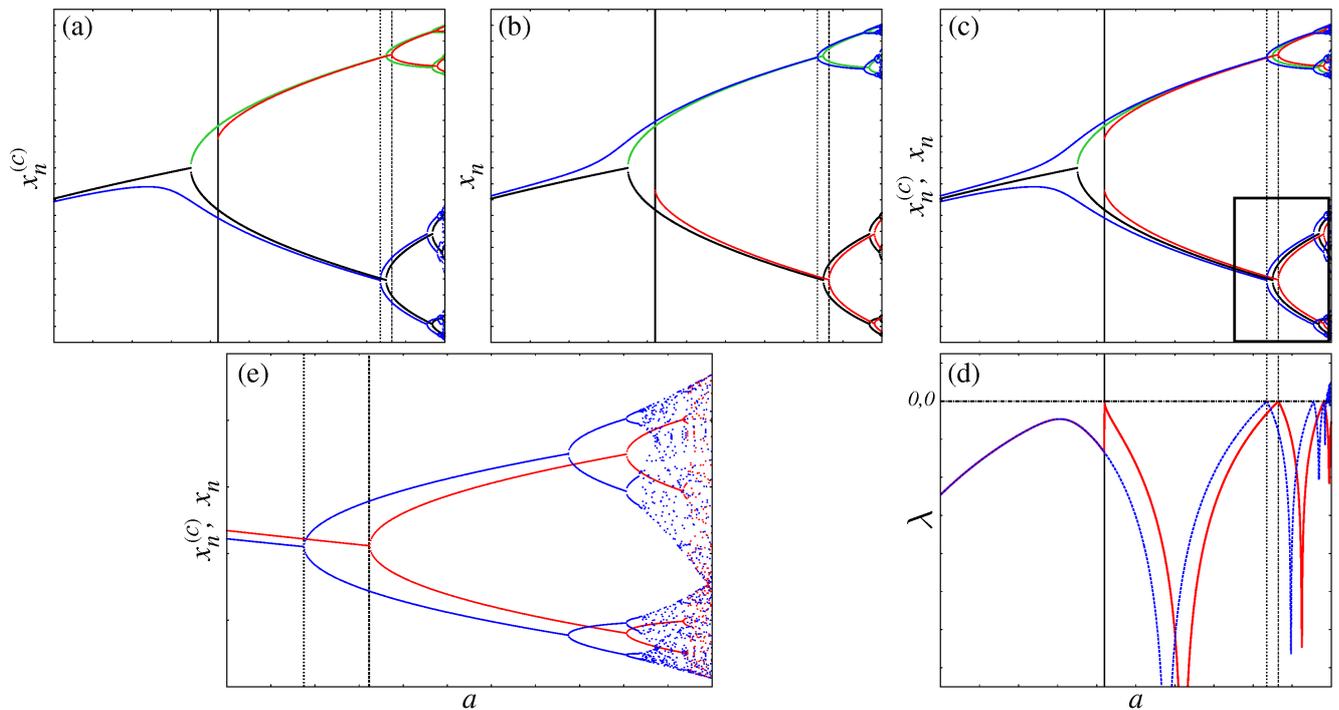}
  \caption{(Color online) {Bifurcation diagram for the {composition 
  of} MQMs {with $k=2$}. The black and green curves represent different attractors 
  for the case $F=0$ while  red and blue curves represent different attractors for the 
  case $F=7\times 10^{-3}$.  In (a) the composed map $x^{(c)}_n$ (see text) is 
  plotted, in (b) only the intermediate  points $x_n$ are plotted and in 
  (c) $x^{(c)}_n$ and $x_n$ are displayed together, all  these cases for 
  the interval $(a_{\mbox{\tiny min}}, a_{\mbox{\tiny max}})=(0.4,1.4)$  and 
  $(x_{\mbox{\tiny min}}, x_{\mbox{\tiny max}})=(-0.6,1.5)$. In (d), the Lyapunov 
  exponent for the composed map $x^{(c)}_n$ is shown for same interval of 
  $a$ and $(\lambda_{\mbox{\tiny min}}, \lambda_{\mbox{\tiny max}})=(-3.0,0.5)$, 
  demonstrating the occurrence of bifurcations ($\lambda=0$) for the values 
  of $a$ indicated by the vertical lines. In (e) we show the magnification 
  of the black box in (c) to evidence the shifted and duplicated bifurcation 
  diagram when $x^{(c)}_n$ and $x_n$ are considered for $F\neq 0$.}}
  \label{logi}
\end{figure}
\end{widetext}

We start with the most simple case of duplication, where the time 
parameter variation has a period $k=2$. In such case the modified
quadratic map (MQM) is defined as 
\begin{equation}
\label{f2}
x_{n+1} = a - x_n^2 + F\,(-1)^n,
\end{equation}
with $F$ being the intensity which changes the dynamics
for  each  iteration and the signal  $(+)$ is used for $n$  
even and $(-)$ for $n$ odd. The {\it composed} map 
{
\begin{equation}
\label{f2c}
x^{(c)}_{n+2} = a - [a - (x_n^{(c)})^2+F]^2 - F,
\end{equation}}
is a composition of two MQMs with  alternating values of $F$, 
namely $+F,-F,+F,-F,\ldots$. In fact, {in this case the intermediate
dynamics} corresponds to a quadratic map
$x^{\prime}_{n+1} = a^{\prime}_n - x^{\prime^2}_n$ with time dependent
parameter $a^{\prime}_n=a+F\,(-1)^n$. A per-$1^c$
orbit from the composed map obeys {$x^{(c)}_2=a - x_1^2 - F=x_0^{(c)}$ 
with $x_1=a-[x_0^{(c)}]^2+F$ and $x_2^{(c)}$} is the fixed point from the 
composed map. The {\it intermediate} point $x_1$ is not a fixed point 
from the composed map but it is necessary to realize the connection 
{$x_0^{(c)}\to x_1\to x^{(c)}_2=x_0^{(c)}$.} 

{It is interesting to observe the bifurcation diagram 
for the composed map (\ref{f2c}) with $F=0$, plotted in 
Figs.~\ref{logi}(a)-(c) (black and green curves). The shape of
this bifurcation  diagram is identical to the bifurcation diagram for 
the usual QM, Eq. (\ref{f2}) with $F=0$, with one essential difference 
explained next. It is known that for the usual QM a period-$1$ (shortly written 
as per-$1$) is born at $a_1=-0.25$, and period doubling bifurcations 
(PDBs) $1\to 2$ and $2\to 4$ occur respectively at $a_{12}=0.75$ and 
$a_{24}=1.25$. For the composed map, inside the interval $[a_{12},a_{24}]$,
instead a per-$2$ orbit, we have {\it two} per-$1^c$ orbits, one drawn
in black and the other one in green. Thus, to obtain the bifurcation
diagram for the composed map it is necessary to use more ICs.}
%
\begin{figure}[!t]
  \centering
  \includegraphics*[width=1.0\columnwidth]{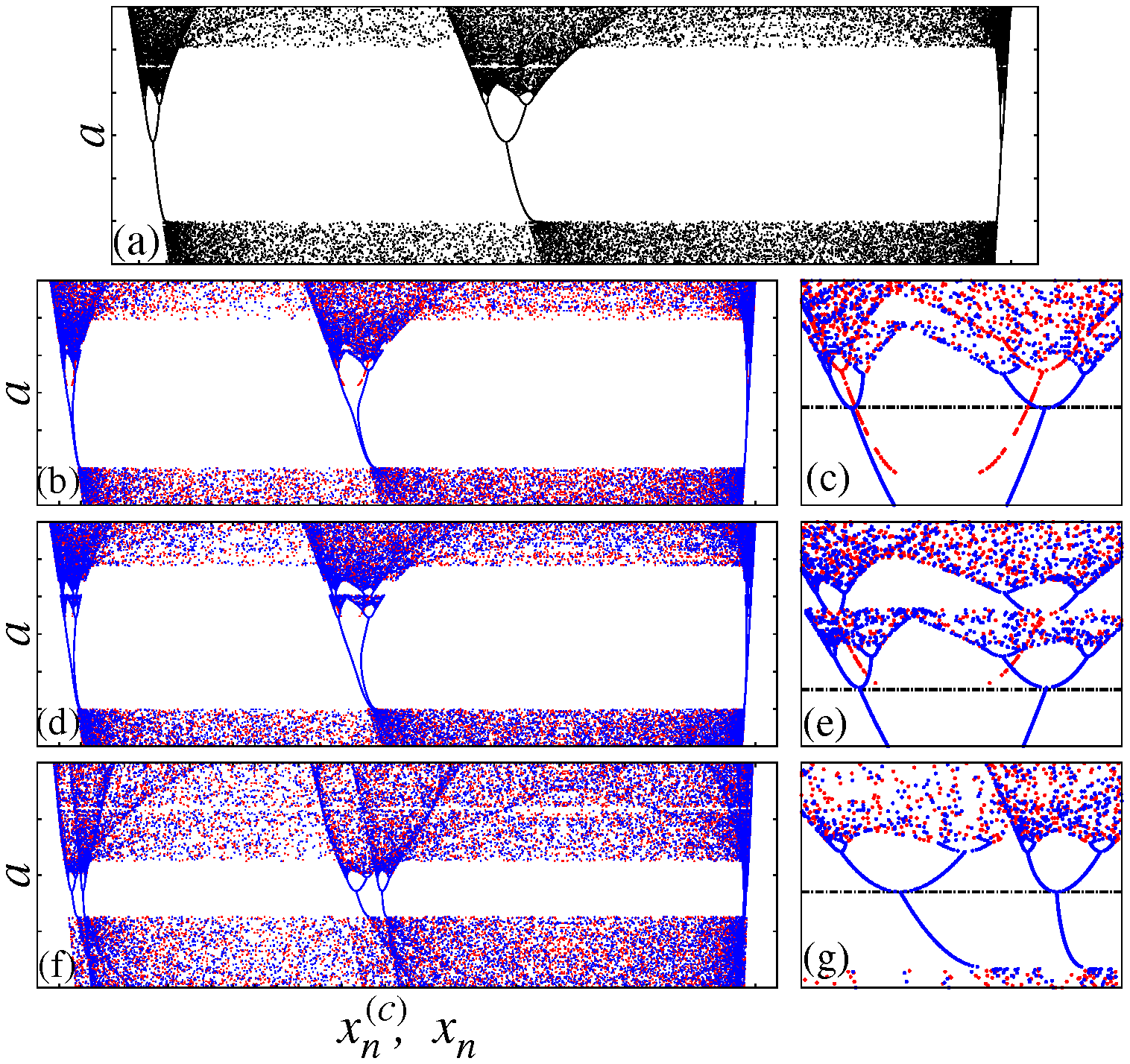}
  \caption{(Color online) {Bifurcation diagram for the composition of 
     MQMs with $k=2$ (sequence $+F,-F,+F,-F,\ldots$) {plotting the composed 
     map $x^{(c)}_n$ and intermediate points $x_n$}, showing a per-3 window
     for (a) $F=0$, (b) and (c) $F=3\times 10^{-4}$, (d) and (e) $F=7\times 
     10^{-4}$, (f) and (g) $F=8\times 10^{-3}$. The black horizontal lines 
     in (c), (e) and (g) indicate the PDB $1 \to 2$ that occurs at the same 
     value of $a$ for each branch of the per-3 window of the $F=0$ case. In 
     (a), (b), (d) and (f) the interval of $x$ and $a$ used are $(x_{\mbox{\tiny
      min}},x_{\mbox{\tiny max}})=(-1.50,1.90)$ and
      $(a_{\mbox{\tiny min}},a_{\mbox{\tiny max}})=(1.74,1.80)$.}} 
  \label{k2p3}
\end{figure}

{Now we use  $F=7\times 10^{-3}$ to analyze the bifurcation diagram 
for the composed map in comparison with the QM. This is shown in 
Fig.~\ref{logi}(a), blue curve for the IC used to obtain the black curve 
from the $F=0$ case, and red curve for the IC related to the green curve 
from the $F=0$ case.} Thus the per-$1$ from the QM is transformed in a per-$1^c_1$ 
(blue curve) orbit from the composed map. This orbit does not suffer a PDB at 
$a_{12}=0.75$ anymore, but at $a_{12}\sim1.23$. 
The PDB at $a_{12}=0.75$ becomes {\it forbidden} for the composed map
and the per-$1^c_1$ avoids the bifurcation point at 
$a_{12}$. Instead, a new per$1^c_2$ (red curve) orbit is born at {$a=0.820$} 
{(see the black-continuous line in Figs.~\ref{logi}(a)-(c))}. Actually, it 
is a structural change in the dynamics when compared to the $F=0$
case. The per-$1^c_1$ and per-$1^c_2$ orbits are independent 
{(different ICs)}, have distinct stabilities and suffer a PDB $1^c\to 2^c$ 
at {\it distinct} values of $a$, namely per-$1^c_1$ at $a=1.235$ and per-$1^c_2$ at
$a=1.265$, {as represented in Fig.~\ref{logi} by the black dotted and 
black dashed lines, respectively}. This is the explanation for the shift 
(to smaller and larger values of $a$) in the bifurcation diagram and it is 
essentially related to the prohibition of a PDB at $a_{12}$ due to $F\ne0$. 
{Figure \ref{logi}(b) shows the bifurcation diagram for the intermediate 
dynamics $x_n$. Essentially it has a similar (mirror symmetric) behavior than 
those shown in Fig.~\ref{logi}(a).} Thus we have generated, including the 
intermediate points, two independent and shifted bifurcation diagrams as shown 
in Fig.~\ref{logi}(c) and magnified in Fig.~\ref{logi}(e). The Lyapunov exponent 
from all orbital points for the composed map is shown in
Fig.~\ref{logi}(d). It is worth to mention that if we use the sequence 
$-F,+F,-F,+F,\ldots$ instead   
$+F,-F,+F,-F\ldots$, intermediate and orbital points are switched, keeping
exactly the same properties discussed above. 

At this {stage}, a natural question that arises is related 
to how the composition of one-dimensional maps perturbs the position of 
fixed {points} in phase and parameter spaces. In order to clarify this 
{issue}, we derive analytical expressions for the orbits 
of the composed map for the shifted bifurcation parameter.
For the born of per-$1^c_1$ and per-$1^c_2$ orbits via saddle-node 
bifurcation the expression is  
\begin{eqnarray}
  \nonumber
  W_{1^c}^{(k=2)}(a,F) = &-&256F^4 + (512a^2+1536a+288)F^2 \\
  &-& W_1(a)\,W_{1\to 2}(a)=0,
\label{w1}
\end{eqnarray}
while for the PDB $1^c\to2^c$ it is given by
\begin{eqnarray}
  \nonumber
  W_{1^c\to2^c}^{(k=2)}(a,F) = &-& 256F^4 + (512a^2+1536a+160)F^2 \\
  &-& W_{2\to4}(a)=0.
\label{w12}
\end{eqnarray}
Equations (\ref{w1}) and (\ref{w12}) are written as functions of the 
boundary conditions in parameter space 
{from the usual QM}, namely $W_1(a)=(4a+1)=0 \rightarrow (a=-1/4)$ 
which gives the saddle-node bifurcation, $W_{1\to 2}(a)=(4a-3)^3=0 \rightarrow 
(a=3/4)$ the PDB $1\to2$ and $W_{2\to4}(a)=(16a^2+8a+5)(4a-5)^2=0
\rightarrow (a=5/4,-1/4\pm i/2)$ the PDB $2\to4$ (see \cite{gallas95} 
for details how to obtain such boundary conditions in parameter space). Note 
that the {composed map} {\it couples} $W_1(a)$ and 
$W_{1\to 2}(a)$ and the PDB $1\to2$ is now inserted into the expression (\ref{w1}) 
which is for the saddle-node bifurcation of per-$1^c_1$ and per-$1^c_2$ orbits. 
Thus, the PDB from the QM is transformed into a saddle-node 
bifurcation for the {composed map}, which demonstrates analytically 
the prohibition of the PDB $1\to2$. For the case discussed above, with $F=7\times 
10^{-3}$, the solutions of $W_{1^c}^{(k=2)}$ and $W_{1^c\to2^c}^{(k=2)}$ are, 
respectively, $a=-0.249$ (born of per-$1^c_1$, blue curve), $a= 0.820$ (born of 
per-$1^c_2$, red curve) and $a_{12}=1.235, 1.265$, where the PDBs $1^c\to2^c$ of 
per-$1^c_1$ and per-$1^c_2$, respectively, occur. Note that the complex solution 
$-1/4\pm i/2$ from  $W_{2\to4}(a)=0$ is transformed into a real solution leading 
to {\it two} PDBs $1^c\to2^c$. This proves analytically the origin of shifted 
bifurcation diagrams at distinct values of $a_{12}$.

Now we present details for the case $k=2$ applied to a per-3 window
of the QM in Fig.~\ref{k2p3}. Figs.~\ref{k2p3}(b), (d) and (f) show
the bifurcation diagram for increasing values of $F$ and
Figs.~\ref{k2p3}(c),(e) and (g) display corresponding
magnifications. As expected, the per-$3$ orbit is transformed in
a per-$6$ one, {\it i.e.}, a per-3$^c$ orbit of 
the composed map, and the usual PDB $3\to 6$ of the  QM 
becomes prohibited (see Fig.~\ref{k2p3}(c)), as in the per-$1$ case discussed 
above. In this case, the duplication process is only allowed in the region of 
per-6 orbit of the QM, after the PDBs indicated by the horizontal 
lines in Figs.~\ref{k2p3}(c), (e) and (g). Therefore, we have three additional 
shifted and partially superposed bifurcation diagrams which for the largest 
values of $F$ in Fig.~\ref{k2p3}(f), are embedded in the chaotic zone. This 
is the reason of stabilization and suppression of the chaotic motion, {\it i.e.}, 
periodic and chaotic attractor are coexisting in the same set of parameter $a$. 
With these results we conclude that the shift in the bifurcation diagram and the 
duplication of stable regions is only allowed for even periods when $k=2$, while 
odd periods $p$ becomes $p'=2p$ (per-1 becomes per-2 in Fig. 
\ref{logi} and per-3 becomes per-6 in Fig. \ref{k2p3}), but are not shifted.

\subsection{Triplication ($k=3$)}
%
\begin{widetext}
$\quad$
\begin{figure}[!b]
  \centering
  \includegraphics*[width=0.97\columnwidth]{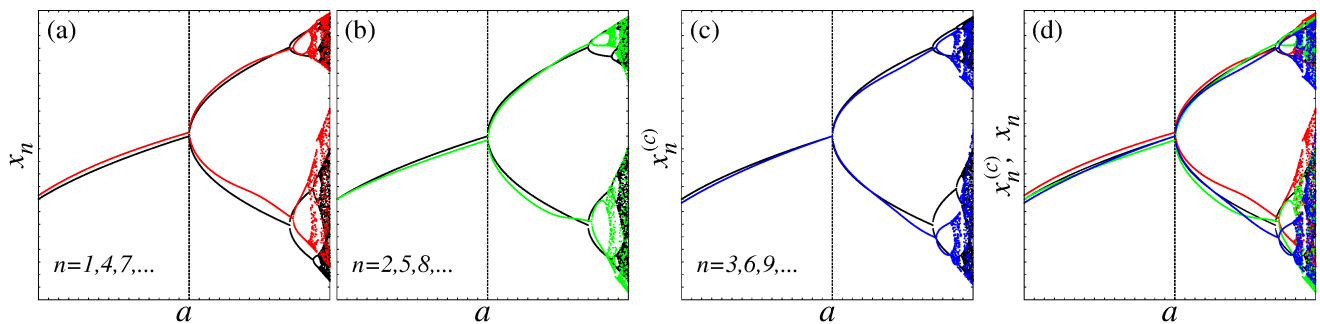}
  \caption{(Color online) {Bifurcation diagrams for a sequence
      $+F,0,-F$ ($k=3$) with $F=3\times 10^{-2}$ (colored dots) using the 
      intermediate points $x_n$ in (a)-(b), the composed map $x^{(c)}_n$ in 
      (c) and both in (d). The black dots represent the bifurcation diagram 
      for the QM in all panels.} The parameter $a$ and variable $x$ range used in 
      all bifurcation diagrams are $(a_{\mbox{\tiny min}},a_{\mbox{\tiny max}})=
      (0.00,1.45)$ and $(x_{\mbox{\tiny min}},x_{\mbox{\tiny max}})=(-0.80,1.50)$, 
      respectively.}  
    \label{tripli1}
\end{figure}
\end{widetext}

To understand more complex behaviours, it is necessary to discuss 
in details the case of triplication. With this purpose 
in mind  we perturb the dynamics of the QM by using the 
$k=3$ sequence $-F,0,+F,-F,0,+F\ldots$.

The per-$1$ orbit from the composed map becomes a composition of three 
orbital points from the intermediate dynamics. The intermediate dynamics 
can be observed in Figs.~\ref{tripli1}(a)-(b) while {the composed map 
is plotted in Fig \ref{tripli1}(c) and} the total dynamics is displayed in 
Fig.~\ref{tripli1}(d). In Figs~(a)-(c) only the iterations 
$n=1,4,7,\ldots$ (for $+F$), $n=2,5,8,\ldots$ (for $0$) and, $n=3,6,9,\ldots$ 
(for $-F$) are plotted in red, green and blue dots, respectively. For 
comparison, all bifurcation diagrams are superposed to the bifurcation diagram 
of the QM, plotted with black dots. In this scenario the PDB is not prohibited 
and the bifurcation points are no shifted as before. Only the phase space is 
composed of more stable points (or periodic attractors). The stable regimes 
in the parameter $a$ remain unaltered.

\begin{widetext}
$\quad$
\begin{figure}[!t]
  \centering
  \includegraphics*[width=0.8\columnwidth]{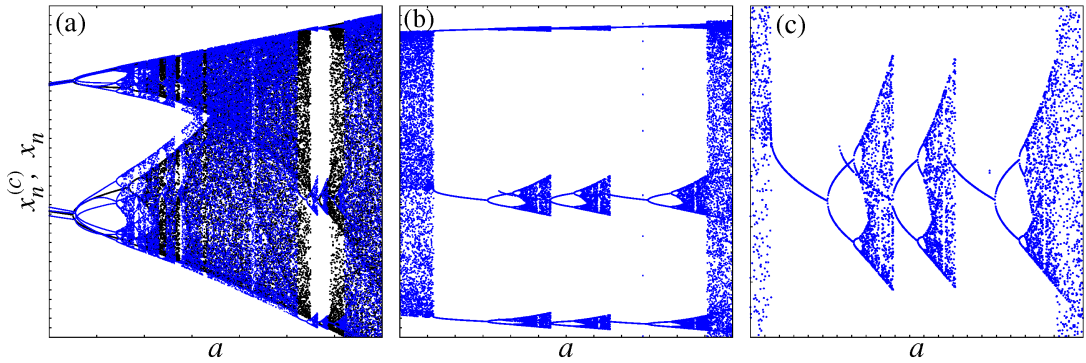}
  \caption{(Color online) {Bifurcation diagram for the composition of MQMs 
    using a sequence $+F,0,-F$ $(k=3)$ which generates the triplication of a 
    per-3 window of the QM. In (a) the QM (black dots), {the composed map 
    $x^{(c)}_n$ and intermediate points $x_n$ (blue dots)} for $F=1.5\times 10^{-3}$ 
    are plotted in the interval of $(a_{\mbox{\tiny min}}, a_{\mbox{\tiny max}})=
    (1.2,1.9)$ and $(x_{\mbox{\tiny min}},x_{\mbox{\tiny max}})=(-1.5,2.0)$, 
    while in (b) only $x^{c}_n$ and $x_n$ are plotted showing the triplication of 
    the per-3 window (or per-1 of the composed map) in the interval of 
    $(a_{\mbox{\tiny min}},a_{\mbox{\tiny max}})=(1.71,1.83)$ and
    $(x_{\mbox{\tiny min}},x_{\mbox{\tiny max}})=(-1.5,2.0)$. Panel  
    (c) is a magnification of small portion of Fig.~(b) for the interval 
    $(a_{\mbox{\tiny min}},a_{\mbox{\tiny max}})=(1.715,1.830)$ 
    and $(x_{\mbox{\tiny min}},x_{\mbox{\tiny max}})=(-0.3,0.3)$ where
    three similar small bifurcation diagrams are generated by the
    MQM.}}
  \label{triplip3}
\end{figure}
\end{widetext}

\begin{figure}[!t]
  \centering
  \includegraphics*[width=0.98\columnwidth]{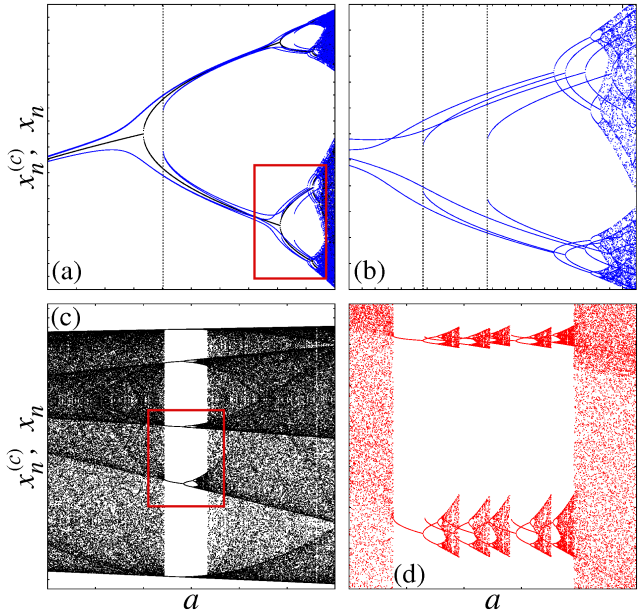}
  \caption{(Color online) Bifurcation diagram for the quadruplication and 
  quintuplication of the QM {plotting $x^{(c)}_n$ and $x_n$}. In (a) the 
  quadruplication is shown for $F=1\times 10^{-2}$ (blue curves) inside the 
  interval $(a_{\mbox{\tiny min}}, a_{\mbox{\tiny max}})= (0.40,1.45)$,
  $(x_{\mbox{\tiny min}}, x_{\mbox{\tiny max}})=(-0.70,1.50)$.  
  For comparison the {bifurcation diagram for the QM} is shown with a black 
  curve. In (b) the magnification of the region inside the red box of (a) is
  plotted and black dotted lines indicate the saddle-node bifurcations
  that create new attractors. In (c) a per-5 window of the QM is 
  plotted inside the interval $(a_{\mbox{\tiny min}}, 
  a_{\mbox{\tiny max}})=(1.60,1.66)$, $(x_{\mbox{\tiny min}},
  x_{\mbox{\tiny max}})=(-1.15,1.90)$. In (d) the quintuplication of
  two orbital points [see red box in (c)] and the corresponding  
  shifted bifurcation diagrams are shown for $F=4\times10^{-3}$.}
  \label{quadri}
\end{figure}

However, shifted bifurcations may occur for other periodic orbits,
namely, periods multiple of 3. To clarify this process a larger 
portion of the bifurcation diagram is magnified in Fig.~\ref{triplip3}(a),  
which includes a per-$3$ window from the QM in black and the composed map in 
blue dots. For smaller values of $a$ there are no shifted bifurcation diagrams, 
as previously shown in Fig.~\ref{tripli1}(d). On the other hand, as $a$ 
increases, some regular windows are already shifted when compared to the 
$F=0$ case. The reason for these shifts are explained using the larger per-$3$ 
window, which are magnified in Figs.~\ref{triplip3}(b) and nicely visualized in 
(c). As observed for the duplication of per-$2$ attractors, in the triplication 
of per-$3$ orbital points, three shifted saddle-node bifurcations
occur, while for the QM only one saddle-node bifurcation 
gives rise to the per-3 window for $a=7/4$. It is important to note that the
per-3 window is enlarge and the period remains the same, while for
other periods $p$, that are not multiple of $k=3$, become $p'=3p$ when
$F\neq 0$. 

\subsection{Quadruplication and quintuplication}

In the present Section the recipe discussed above is extended
to quadruplicate and quintuplicate attractors in phase space to obtain
similar shifted bifurcation diagrams. For the quadruplication and 
quintuplication processes we use the protocols 
$+F,-F/2,+F/2,-F,+F,-F/2,\ldots$ ($k=4$) and $+F,-F/2,0,
+F/2,-F,+F,-F/2,\ldots$ ($k=5$), respectively. For $k=4$, the  
multiplication of stable regions occurs as in the case $k=2$. However 
not only the PDB $1 \to 2$ is prohibited but also $2 \to 4$, giving rise to 
four saddle-node bifurcations, displayed by vertical dotted lines in
Figs.~\ref{quadri}(a) and (b). The case $k=5$ can be compared  
to the case $k=3$ and it is shown in Fig \ref{quadri}(c) ($F=0$) and
(d) ($F\neq 0$).

\section{Conclusions} 
\label{conclusions}

In this work we show that the control of intermediate dynamics  
can be used to enlarge stable domains in phase and parameter spaces of nonlinear 
dynamical systems. We present analytical and numerical results for the 
specific case of the composition of quadratic maps with distinct parameters.  
In general, performing many simulations we observed that  the sign of
the perturbation  must change for each composition of the map. In
addition, the  $k$-compositions generate $k$ attractors and
$k$-shifted independent bifurcations diagrams when
$\omega\in\mathbb{Z}$, where $\omega=p/k$ and $p$ being the period of
the orbits from the QM. For $F\ne 0$ the original period 
of the orbit remains equal. When $\omega\notin\mathbb{Z}$ we still generate
$k$-attractors, but the  bifurcation diagrams are not shifted
anymore. In this case the period of the orbits {obeys} $p^{\prime}=pk$
since it counts also the intermediate orbits.

The ability to control the intermediate dynamics in realistic
situations has relevant implications: (i)  multiple composition of
maps lead to the appearance of multiple shifted  bifurcation
diagrams. Consequently occurs a considerable  enlargement of the
stable domains in phase  and parameter spaces.  This is crucial for
the survival of  the desired dynamics under noise and temperature
effects, which usually destroy periodic motion
\cite{alan13-1,cesar17};  (ii) the number $k$ of modified maps  used
in the composition determinates the period-$p$ of multiplied
bifurcation diagrams since the necessary condition is that $\omega \in
\mathbb{Z}$, with $\omega=p/k$. 

The easy way of generating and moving shifted bifurcation diagrams by
controlling the intermediate dynamics of composed maps is definitely
the remarkable contribution of the present work. {Consequences of
shifted bifurcations in two dimensional systems were analyzed recently 
by multiplying isoperiodic stable structures in the  parameter space of
the Hénon map \cite{rafael17-2}}. Future contributions intend to verify 
to which extend such multiplication of stable motion can be realized in 
the parameter space of classical \cite{alan11-1,alan11-2} and quantum 
ratchets \cite{alan15}.

\acknowledgments{R.M.S. thanks CAPES (Brazil) and C.M. and M.W.B. thank 
CNPq (Brazil) for financial support. C.M. also thanks FAPESC (Brazil)
for financial support. The  authors  also  acknowledge computational 
support from Professor Carlos M.~de Carvalho at LFTC-DFis-UFPR.}



\end{document}